\documentclass[twocolumn]{aastex631}

\usepackage{amsmath}
\usepackage{gensymb}
\usepackage{xspace}

\newcommand\tierrasenddate{2025~March~20}

\newcommand{\rmfit}{\texttt{rmfit}\space}

\newcommand{\Rstar}{\ensuremath{R_{\star}}\xspace} 
\newcommand{\Mstar}{\ensuremath{M_{\star}}\xspace}
\newcommand{\Rjup}{\ensuremath{R_\mathrm{J}}\xspace} 
\newcommand{\Mjup}{\ensuremath{M_\mathrm{J}}\xspace}

\newcommand{\Rsun}{\ensuremath{R_\odot}\xspace} 

\newcommand{\Rp}{\ensuremath{R_\mathrm{p}}\xspace}
\newcommand{\Mp}{\ensuremath{M_\mathrm{p}}\xspace}

\newcommand{\vsini}{\ensuremath{v\sin{i_\star}}\xspace}
\newcommand{\veqsini}{\ensuremath{v_\mathrm{eq}\sin{i_\star}}\xspace}

\newcommand{\Prot}{\ensuremath{P_\mathrm{rot}}\xspace}
\newcommand{\istar}{\ensuremath{i_{\star}}\xspace}
\newcommand{\iorb}{\ensuremath{i_\mathrm{orb}}\xspace}

\newcommand{\ms}{\ensuremath{\mathrm{m}\,\mathrm{s}^{-1}}\xspace}

\newcommand{\kms}{\ensuremath{\mathrm{km}\,\mathrm{s}^{-1}}\xspace}

\newcommand{\prot}{\ensuremath{23.47\pm0.29}}
\newcommand{\measuredL}{\ensuremath{43.68^{+11.67}_{-8.97}}}
\newcommand{\measuredSig}{\ensuremath{4.9}}
\newcommand{\measuredSigNtoN}{\ensuremath{3.0}}

\defcitealias{Yee2023}{Y23}

\newcommand{\FitLambda}[1]{\ensuremath{{7#1}^{+10#1}_{-11#1}}}
\newcommand{\FitPsi}[1]{\ensuremath{{15.6#1}^{+7.7#1}_{-7.3#1}}}
\newcommand{\FitVsini}{\ensuremath{1.84^{+0.05}_{-0.06}}}
\newcommand{\FitIstar}[1]{\ensuremath{90#1 \pm 13#1}}

\received{April 9, 2025}
\revised{May 2, 2025}
\accepted{May 5, 2025}

\shorttitle{True Obliquity of TOI-2364}
\shortauthors{Tamburo et al.}
\graphicspath{{./}{figures/}}

\begin{document}

\title{The True Stellar Obliquity of a Sub-Saturn Planet from the \textit{Tierras} Observatory and KPF} 


\author[0000-0003-2171-5083]{Patrick Tamburo}
\affiliation{Center for Astrophysics $\vert$ Harvard \& Smithsonian, 60 Garden Street, Cambridge, MA 02138, USA}

\author[0000-0001-7961-3907]{Samuel W. Yee}\altaffiliation{51 Pegasi b Fellow}
\affiliation{Center for Astrophysics $\vert$ Harvard \& Smithsonian, 60 Garden Street, Cambridge, MA 02138, USA}

\author[0000-0003-1361-985X]{Juliana Garc\'ia-Mej\'ia}\altaffiliation{51 Pegasi b Fellow}
\affiliation{Center for Astrophysics $\vert$ Harvard \& Smithsonian, 60 Garden Street, Cambridge, MA 02138, USA}
\affiliation{Kavli Institute for Astrophysics and Space Research, Massachusetts Institute of Technology, Cambridge, MA 02139, USA}

\author[0000-0001-7409-5688]{Gudmundur Stef{\'a}nsson}
\affiliation{Anton Pannekoek Institute for Astronomy, University of Amsterdam, 904 Science Park, Amsterdam, 1098 XH}

\author[0000-0002-9003-484X]{David Charbonneau}
\affiliation{Center for Astrophysics $\vert$ Harvard \& Smithsonian, 60 Garden Street, Cambridge, MA 02138, USA}

\author[0000-0001-6637-5401]{Allyson~Bieryla}
\affiliation{Center for Astrophysics $\vert$ Harvard \& Smithsonian, 60 Garden Street, Cambridge, MA 02138, USA}

\author[0000-0001-8638-0320]{Andrew~W.~Howard}
\affiliation{Department of Astronomy, California Institute of Technology, Pasadena, CA 91125, USA}
\author[0000-0002-0531-1073]{Howard~Isaacson}
\affiliation{Department of Astronomy,  University of California Berkeley, Berkeley, CA 94720, USA}

\author[0000-0003-3504-5316]{Benjamin~J.~Fulton}
\affiliation{NASA Exoplanet Science Institute / Caltech-IPAC, Pasadena, CA 91125, USA}

\author[0000-0002-5812-3236]{Aaron Householder}
\affiliation{Kavli Institute for Astrophysics and Space Research, Massachusetts Institute of Technology, Cambridge, MA 02139, USA}
\affiliation{Department of Earth, Atmospheric and Planetary Sciences, Massachusetts Institute of Technology, Cambridge, MA 02139, USA}


\begin{abstract}
We measure the true obliquity of TOI-2364, a K dwarf with a sub-Saturn-mass ($M_p = 0.18\,M_J$) transiting planet on the upper edge of the hot Neptune desert. We used new Rossiter-McLaughlin observations gathered with the Keck Planet Finder to measure the sky-projected obliquity $\lambda = \FitLambda{^\circ}$. Combined with a stellar rotation period of \prot~days measured with photometry from the \textit{Tierras} Observatory, this yields a stellar inclination of $\FitIstar{^\circ}$ and a true obliquity $\psi = \FitPsi{^\circ}$, indicating that the planet's orbit is well aligned with the rotation axis of its host star. The determination of $\psi$ is important for investigating a potential bimodality in the orbits of short-period sub-Saturns around cool stars, which tend to be either aligned with or perpendicular to their host stars' spin axes.

\end{abstract}

\section{Introduction}\label{sec:intro}

The planets in our solar system orbit the Sun with low mutual inclinations relative to each other and to the Sun’s spin axis, consistent with formation in a protoplanetary disk.
The discovery of exoplanets on orbits significantly misaligned with the stellar equators of their host stars \citep{Hebrard2008,Winn2009b} suggests that either the disks themselves can be misaligned, or that dynamical processes can torque planetary orbits out of alignment with the stellar spin axis. The majority of measurements of spin-orbit alignments for exoplanetary systems have been made using the Rossiter-McLaughlin (RM) technique (see, e.g. review by \citealt{Albrecht2022}).
However, the RM and other transit-based techniques constrain only the sky-projected angle between the stellar spin axis and the orbit normal $\lambda$, and not the three-dimensional obliquity, $\psi$. 

The missing dimension may obscure population-level trends that provide important insights into the mechanisms driving spin-orbit alignment. \citet{Albrecht2021} found that when examining the distribution of true obliquities, there was an apparent preference for $\psi$ to cluster at 90$^\circ$, a feature that was less apparent in the distribution of projected obliquities.
This preponderance of polar orbits may be a phenomenon particular to sub-Saturn-mass ($\Mp \lesssim 0.3\,\Mjup$) planets \citep{Attia2023,Knudstrup2024,Espinoza-Retamal2024,Handley2024}, consistent with some dynamical theories for misaligning a planet with its star \citep[e.g,][]{Petrovich2020}.
However, the significance of the pile-up of polar orbits is still in question. Using the full sample of sky-projected obliquities, \citet{Siegel2023} and \citet{Dong2023} found weakened evidence for this peak, and suggested that the set of $\psi$ measurements may have been biased in some way that produces the pile-up.
A larger sample of planets with 3D obliquity measurements may help resolve this tension.

One way to obtain $\psi$ is to determine the stellar inclination \istar by combining measurements of $v\sin \istar$ (from spectroscopic or RM measurements) with the stellar rotation period, $P_\mathrm{rot}$: 

\begin{equation}
    \istar = \sin^{-1} \left(\frac{\vsini}{v}\right) = \sin^{-1} \left(\frac{\vsini}{2\pi R_\star/P_{\mathrm{rot}}}\right) 
\end{equation}

\noindent where $R_\star$ is the stellar radius \citep[note, however, that $v$ and $v\sin i$ are correlated, and this must be accounted for properly following][]{Masuda2020}. However, for field-age main-sequence stars, with characteristic rotation periods of weeks and amplitudes of less than a percent,  $P_\mathrm{rot}$ is not easily measured. The Kepler and K2 missions \citep{Bo10a, Ho14} measured $P_\mathrm{rot}$ for tens of thousands  of main-sequence stars \citep[e.g.,][]{Nielsen2013, McQuillan2014, Reinhold2020, Gordon2021}, but over a small area of the sky. The Transiting Exoplanet Survey Satellite \citep[TESS;][]{Ri15} is providing high-precision photometry for bright stars across most of the sky, but these data are generally poorly suited for measuring rotation periods of field-age Sun-like stars, as a single TESS sector is about 27 days, typically limiting rotation period measurements from TESS to about half a sector length \citep[e.g.,][]{Holcomb2022}\footnote{A notable exception is stars in the TESS continuous viewing zones \citep{Claytor2024}.}. Ground-based all-sky surveys like the All-Sky Automated Survey for Supernovae \citep[ASAS-SN;][]{Kochanek2017} and the Zwicky Transient Facility \citep[ZTF;][]{Bellm2018} collect data over baselines of years, but are subject to systematic effects from the wavelength-dependent and time-variable transmission that afflicts broadband photometry, which has historically limited the night-to-night precision of ground-based light curves to several parts-per-thousand \cite[ppt; e.g.,][]{Lockwood2007}.

The \textit{Tierras} Observatory is a 1.3-m telescope at Fred Lawrence Whipple Observatory atop Mt. Hopkins in Arizona. It is equipped with a photometer that uses a narrow-band near-infrared filter ($\lambda_0 = 853.5$~nm; $\mathrm{FWHM} = 40.2$~nm) that was designed to limit photometric errors due to precipitable water vapor lines \citep{GaMe20}. In practice, we have found that we are able to maintain night-to-night stability down to 0.5~ppt over baselines of months on bright, non-variable stars, and we are working toward a night-to-night precision goal of 0.25~ppt. One of our science goals is to use \textit{Tierras} to measure $P_\mathrm{rot}$ for a sample of main-sequence stars with planets that may be misaligned based on measurements of $\lambda$.

In this paper, we present the measurement of the true obliquity $\psi$ for TOI-2364, derived from a stellar rotation period measurement from \textit{Tierras} and the detection of the RM effect with the Keck Planet Finder (KPF). 
TOI-2364 is an early K dwarf ($T_\mathrm{eff} = 5300$~K, $M_* = 0.95~M_\odot$, $R_*=0.87~R_\odot$) that hosts a transiting hot sub-Saturn ($\Rp = 0.77\,\Rjup$, $\Mp = 0.18\,\Mjup$) with a 4-day orbital period discovered by TESS (\citealt{Yee2023}; hereafter \citetalias{Yee2023}).

As a low-mass planet around a cool star, TOI-2364\,b belongs to the subpopulation that may demonstrate a preference for polar orbits, for which true obliquity measurements are particularly valuable.

In Section~\ref{sec:obs}, we describe the observations. We detail our analysis of the data in Section~\ref{sec:analysis}. We interpret our measurement of $\psi$ for TOI-2364 in Section~\ref{sec:discussion}. 

\section{Observations}\label{sec:obs}
\subsection{KPF}\label{sec:kpf}
We observed a spectroscopic transit of TOI-2364\,b on UT 22 Oct 2023, using the newly commissioned Keck Planet Finder spectrograph (KPF; \citealt{Gibson2024}) on the Keck-I telescope.
We obtained 31 exposures of 480s each in standard readout mode, over a total observation time of 4.5 hours, covering the full transit of the planet and roughly 2 hours of post-transit baseline.

We reduced the data with the standard KPF Data Reduction Pipeline (KPF-DRP; \citealt{Gibson2020}).\footnote{\url{https://github.com/Keck-DataReductionPipelines/KPF-Pipeline}}
We extracted precise radial-velocities (RVs) by cross-correlating the observed spectra with a weighted line mask derived for G9 stars and used in the ESPRESSO pipeline \citep{Ba96,Pepe2002}.
We derived cross-correlation functions (CCFs) on an order-by-order basis and created a total CCF by summing the CCFs weighted by the expected stellar flux in each order.
We then determined the RV and associated uncertainty for each observation from a Gaussian fit to the total CCF.

\subsection{KeplerCam Photometry}\label{sec:keplercam}
We also obtained photometric observations of TOI-2364 from the KeplerCam CCD on the 1.2m telescope at the Fred Lawrence Whipple Observatory (FLWO), catching the transit ingress contemporaneous with the KPF observations on UT~22~Oct~2023.
KeplerCam uses a $4096\times4096$ Fairchild CCD 486 detector with a field-of-view of $23\farcm1\times23\farcm1$ and an image scale of 0.672$"/$pix when binned by 2. We made observations in the Sloan $i^\prime$ filter with an exposure time of 18~seconds. We performed differential aperture photometry and extracted a light curve for TOI-2364 using AstroImageJ \citep{Co17}.

\subsection{$\mathrm{Tierras}$}\label{sec:tierras}

We observed TOI-2364 with \textit{Tierras} from UT~2024~October~4 to \tierrasenddate. We took observations at a 30-s exposure time and typically performed 2--4 visits to the field over the course of each observing night, each with a duration of about 5 minutes. 

We created a light curve for TOI-2364 using a custom photometric pipeline that we developed for \textit{Tierras}. \textit{Tierras} employs a $4K\times4K$ CCD with a plate scale of 0.43$"/$pix operated in frame-transfer mode, resulting in an on-sky footprint of 0.24$^\circ \times 0.49^\circ$ (R.A.$\times$decl.). We performed photometry and made light curves for all sources that were present in every image out to a Gaia $G_\mathrm{RP}$ magnitude limit of 17, of which there were 388. We used circular apertures with radii ranging from 5--20 pixels (2.2"--8.6"). We measured the local background for each source in each image using circular annuli with inner radii of 35 pixels and outer radii of 55 pixels. The background was taken to be the sigma-clipped mean of the pixels within the annulus with a threshold of 2$\sigma$. We calculated the expected  uncertainty on the target's normalized photometry using the photon noise contribution from the target and sky background, dark current, and read noise.

For each source, we created an artificial light curve (ALC) as the weighted sum of the fluxes from all the other stars. We generated weights following the procedure described in \citet{Tamburo2022a}, with the additional step of totally de-weighting especially noisy stars, as determined by the ratio of their measured standard deviations to their calculated uncertainties, propagating the uncertainties from the ALC correction and adding an estimate of the scintillation noise \citep[e.g.,][]{Stefansson2017}. The aperture size that minimized the scatter on nightly timescales was chosen as the best light curve for each source. For TOI-2364, the light curve used an aperture radius of 8 pixels (3.4$"$). 

We obtained a total of 1856 exposures of TOI-2364 with \textit{Tierras}. For our rotation period analysis (see Section~\ref{subsec:prot}, we excluded 64 in-transit points using the linear ephemeris obtained in Section~\ref{subsec:rm_analysis}. We excluded an additional 538 low-quality points, which we define as having a median FWHM seeing greater than $4"$, a normalized ALC flux less than 0.9, $x$ or $y$ pointing deviations greater than 20 pixels, WCS solutions with an RMS greater than 0.215$"$ (half a \textit{Tierras} pixel), or sky backgrounds greater than 7~ADU/pix/s. We sigma clipped the light curve with a threshold of 4$\sigma$ both on individual nights and globally, excluding an additional 5 points and leaving a total of 1249 photometric data points. 

\subsection{TESS}\label{sec:tess}
TOI-2364 was observed by TESS with in Sectors 6, 33, and 87, with cadences of 30~minutes, 10~minutes, and 2~minutes, respectively. We obtained the Science Processing Operations Center \citep[SPOC;][]{Jenkins2016} light curves for the target and used the Pre-search Data Conditioning Simple Aperture Photometry (PDCSAP) flux in our analysis.
For our analysis of the RM effect, we used only the subset of data containing the planetary transits.
We detrended the light curves by fitting basis splines to the full out-of-transit flux for each sector using the \textit{Keplerspline} code \citep{Keplerspline_Vanderburg2014,Keplerspline_Shallue2018}, before extracting the data corresponding to the planetary transits and baseline flux corresponding to a single transit duration before and after each event.

\subsection{Zwicky Transient Facility}\label{sec:ztf}
We queried the Zwicky Transient Facility \citep[ZTF;][]{Bellm2018} for data on TOI-2364. We downloaded the ZTF \textit{g}-band light curve of the target, which consisted of 312 measurements taken at a 30-s exposure time over a 7-year baseline starting in 2018. Within a season, the median cadence of the ZTF light curve was 3.0~days. We did a 4$\sigma$ clipping of the light curve, removing 4 outliers.

\subsection{All-Sky Automated Survey for Supernovae}\label{sec:asas-sn}
We obtained the All-Sky Automated Survey for Supernovae \citep[ASAS-SN;][]{Shappee2014, Kochanek2017} $V$-band light curve of TOI-2364. These data consisted of 235 measurements taken over five seasons starting in 2014. Within a season, the median cadence of the ASAS-SN light curve was 3.0~days. No outliers were removed from the light curve with a 4$\sigma$ clipping.

\section{Analysis}\label{sec:analysis}

\subsection{Rotation Period Measurement}\label{subsec:prot}
\begin{figure*}
    \centering
    \includegraphics[width=\textwidth]{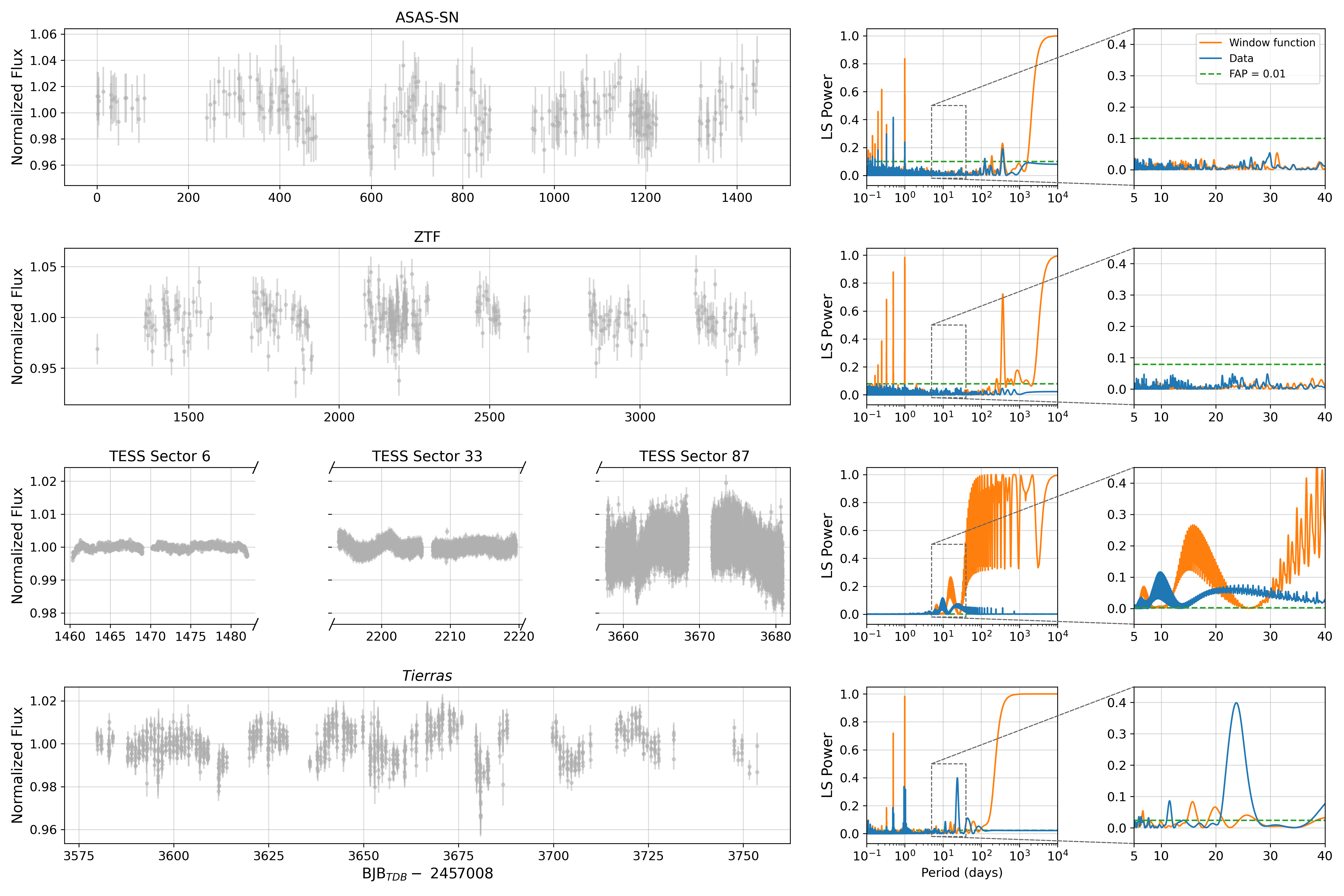}
    \caption{Photometric data used for our initial rotation period search. Each row shows a different data set. Note the different x and y scales for the light curves. LS periodograms of each data set are shown in blue to the right of their respective light curves, with the window function of the data shown in orange. The FAP = 0.01 level is indicated with a dashed green line. We also show a zoom-in on the periodograms from 5--40 days. A significant, narrow peak is detected in the \textit{Tierras} data at 23.5~days.}
    \label{fig:periodograms}

\end{figure*}

We used the \texttt{astropy} \citep{astropy:2013, astropy:2018, astropy:2022} implementation of the Lomb-Scargle (LS) periodogram to search for periodic signals in the TESS, \textit{Tierras}, ZTF, and ASAS-SN datasets described in Section~\ref{sec:obs}. We computed the LS power of the data over a period grid from 0.1--10,000 days. We constructed the window functions of the data over the same period grid. We evaluated the 1\% false-alarm probability (FAP) level for each periodogram using the upper-bound method described in \citet{Baluev2008}. We show the periodograms in Figure~\ref{fig:periodograms}.   

There is a significant single peak in the \textit{Tierras} periodogram at roughly 23.5~days. There is significant power at this period in the TESS periodogram as well, though it is far less localized, consistent with the fact that 23.5~days is near the duration of a TESS sector. There is not a significant peak at 23.5 days in either the ASAS-SN or ZTF data sets. Note that the significant peaks in the ASAS-SN periodogram longer than 100 days are the one-year alias of the data and its third harmonic. We performed an injection and recovery test on the ASAS-SN and ZTF data, injecting sine waves with periods of 23.5~days, random phases, and random amplitudes from 0.001\textendash0.1. We considered an injected signal detected if the highest peak of the resulting Lomb-Scargle periodogram was within 1\% of 23.5~days. For ZTF, we found a detection fraction of about 0.50 for amplitudes around 1\%, the approximate maximum amplitude in the $Tierras$ light curve. For ASAS-SN, we found that no injected signals were recovered with amplitudes below 1.6\%. Thus, the non-detection of a 1\%, 23.5-day signal in the ASAS-SN and ZTF light curves is consistent with the data. 

We used a Gaussian process (GP) model to determine our best estimate of $P_\mathrm{rot}$ and its uncertainty from the \textit{Tierras} data. We modeled the covariance of the data using the following kernel function as implemented in \texttt{celerite} \citep{ForemanMackey2017}: 

\begin{equation}
    \kappa(\tau) = \frac{B}{2+C} e^{-\tau/L} \left[ \cos\left( \frac{2 \pi \tau}{P_{\mathrm{rot}}}\right) + (1+C)\right] + \sigma^2\delta_{\mathrm{ij}}
\end{equation}

In this equation, $\tau$ is a $1249\times 1249$ matrix whose elements represent the absolute difference between timestamps, with $\tau_{\mathrm{i},\mathrm{j}} = |x_\mathrm{i} - x_\mathrm{j}|$. The first term is the \texttt{celerite} version of a quasi-periodic (QP) kernel with hyperparameters $B$, $C$, $L$, and $P_\mathrm{rot}$. $B$ controls the amplitude of the covariance, and with our light curve converted to ppt, $B$ has units of $\mathrm{ppt}^2$. $C$ is unitless. $L$ has units of time and represents the exponential decay timescale of the covariance.  $P_\mathrm{rot}$ also has units of time and represents the rotation period of the star. The QP kernel has been shown to be a reliable model for deriving the rotation periods of stars from photometry \citep{Angus2018,Nicholson2022}. For our kernel we added an additional jitter term, which describes the variance $\sigma^2$ by which our calculated uncertainties are under or overestimated. This variance, multiplied by the Kronecker delta function $\delta_\mathrm{ij}$, gets added onto the diagonal of the covariance matrix resulting from the QP kernel.

We used \texttt{emcee} \citep{Fo13} to sample the posteriors of the natural logarithm of the QP-GP's hyperparameters conditioned on \textit{Tierras} data. The priors on the hyperparameters are given in Table~\ref{tab:gp}. All priors were uniform with the exception of the one on $\ln P_\mathrm{rot}$, for which we used a Gaussian whose central value and standard deviation were estimated from the LS periodogram (see Figure~\ref{fig:periodograms}). The prior on $\ln L$ permits exponential decay timescales ranging from 12 days (about half a rotation period) to 150 days (the approximate duration of the \textit{Tierras} light curve).

\begin{deluxetable}{ll} \label{tab:gp}
\tablecaption{Bound on the QP-GP hyperparameters}
\tablehead{\colhead{Parameter} & \colhead{Prior} }
\startdata
$\ln(B/\mathrm{ppt}^2)$ & $\mathcal{U}(-20.0,0.9)$ \\
$\ln(C)$ & $\mathcal{U}(-20.0,20.0)$ \\
$\ln(L/\mathrm{day})$ & $\mathcal{U}(2.5, 5.0)$\\
$\ln(P_\mathrm{rot}/\mathrm{day})$ & $\mathcal{N}(\ln 23.5, \ln 3.0)$\\
$\ln(\sigma^2/\mathrm{ppt}^2)$ & $\mathcal{U}(1.0, 1.6)$\\
\enddata
\end{deluxetable}

We used 100 walkers and ran the MCMC until the number of steps was greater than 100$\times$ the autocorrelation time. We calculated the residuals between the best-fit sample (as determined by the log probability) and the data and tested for correlations between the residuals and airmass, sky background, and FWHM seeing. We found a significant correlation between the residuals and sky background with a Pearson correlation coefficient of  $0.31$ and a corresponding $p$-value of $2\times10^{-29}$. We ran another MCMC fitting a linear function of the sky to the residuals, removed the best-fit sky model from the data and then re-ran the QP-GP fit. We repeated this procedure until the correlation between the residuals and the sky background had a $p$-value greater than 0.01, which took two iterations. 

\begin{figure*}
    \includegraphics[width=\textwidth]{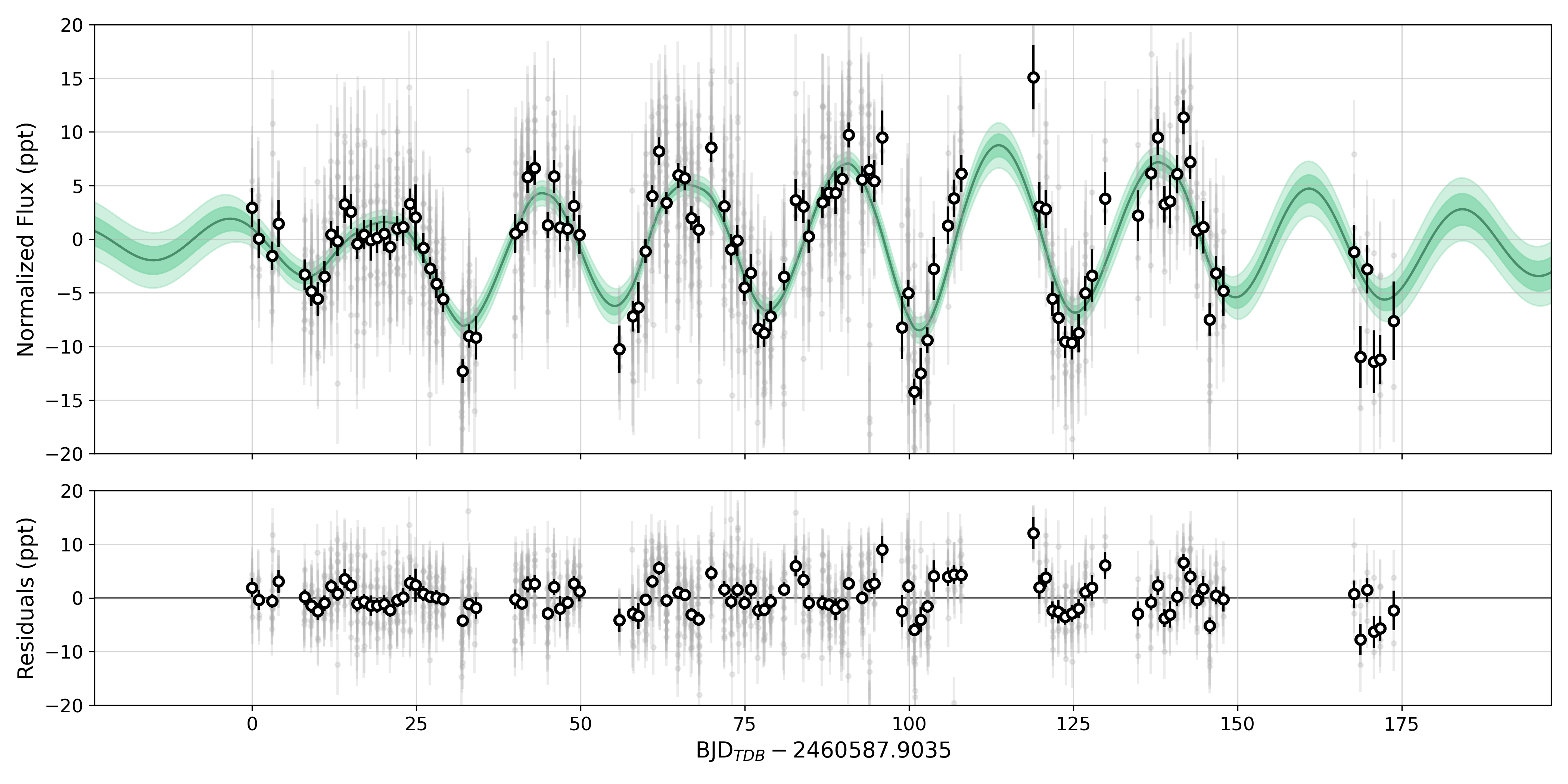}
    \caption{Top: \textit{Tierras} light curve of TOI-2364. Data at the native 30-s cadence are shown in gray, with their errorbars scaled by a factor of 1.66 compared to our calculated photometric uncertainties. White points points with error bars show the median flux measured on each night. The best-fit QP-GP model is shown as a green line, along with its 1- and 2-$\sigma$ uncertainty intervals as shaded regions. Bottom: The residuals from the best-fit QP-GP model.}
    \label{fig:tierras_lc}
\end{figure*}

In Figure~\ref{fig:tierras_lc}, we show the \textit{Tierras} light curve resulting from this procedure, i.e. with the sky background correlation removed. We show the best-fit QP-GP model in green along with its 1 and 2$\sigma$ uncertainty intervals. The uncertainties in this plot have been scaled by a factor of 1.66 in accordance with the jitter term coefficient of the best-fit sample. In the bottom panel we show the residuals of the final fit. The unbinned residuals have a standard deviation of \measuredSig~ppt, while the data binned over each night have a standard deviation of \measuredSigNtoN~ppt. 

We measure $P_{\mathrm{rot}} = $~\prot~days and a covariance decay timescale $L = $~\measuredL~days.

\subsection{RM Modeling} \label{subsec:rm_analysis}

We measured the alignment between the orbit normal of TOI-2364\,b and its host star's spin axis by fitting the KPF RVs to a model of the RM effect.
We used the \rmfit code \citep{rmfit_Stefansson2022}, which models the RM anomaly with the analytic formulae of \citet{Hirano2010}.
\rmfit uses a differential evolution optimization algorithm \citep{PyDE_Storn1997} as implemented in the \texttt{PyDE} code\footnote{\url{https://github.com/hpparvi/PyDE}} to find the global maximum likelihood solution. 
\rmfit then uses the \texttt{emcee} package to sample the posterior probability distributions and derive uncertainties on each of the model parameters.
We used an ensemble of 100 walkers and ran the MCMC chains for a number of steps $> 100\times$ the autocorrelation time.
The MCMC jump parameters, priors, and best-fit values and uncertainties are listed in Table \ref{tab:rm_results}. 

The RM anomaly depends primarily on the sky-projected obliquity $\lambda$ and the projected stellar rotation velocity \vsini, as well as the planet's transit parameters.
\citetalias{Yee2023} reported a measurement of $\vsini = 1.3\pm1.0\,\kms$ based on their analysis of Magellan/PFS spectra using the \texttt{SpecMatch-Syn} code \citep{SpecMatchSynth_Petigura2015}.
However, we did not incorporate this information in our analysis --- spectroscopic \vsini measurements may be biased at such low rotational velocities, as broadening due to macroturbulence, microturbulence, and the instrumental profile become dominant over rotational broadening \citep[e.g.,][]{Masuda2022a}.
Instead, we made use of the rotation period measurement from \textit{Tierras}, from which we can infer the equatorial rotation velocity $v_\mathrm{eq} = 2\pi\Rstar / \Prot$.
We assumed a Gaussian prior of width 0.29~days and centered on the measurement of $\Prot = 23.47$~days from Section \ref{subsec:prot}.
We placed a uniform prior on the distribution of $\cos{i_\star}$, consistent with an isotropic distribution of stellar spin axes.
We then computed
\begin{equation} \label{eq:vsini_eq}
\vsini = 2\pi\Rstar / \Prot \sqrt{1 - \cos^2{i_\star}}
\end{equation}
(neglecting the effects of differential rotation, which we address later).
By including \Prot as one of the parameters in the analysis, we ensured that the posterior distribution inferred for $i_\star$ correctly accounts for the statistical dependence between $v_\mathrm{eq}$ and $\veqsini$ \citep{Masuda2020}.
The true stellar obliquity $\psi$ could then be inferred from the distributions for $i_\star$ and $\lambda$ using
\begin{equation} \label{eq:psi_eq}
    \cos\psi = \sin i_\star\sin i_\mathrm{orb} \cos\lambda + \cos i_\star \cos i_\mathrm{orb}
\end{equation}
where $i_\mathrm{orb}$ is the planet's orbital inclination, which is measured from the photometric transit.

The other input to the \citet{Hirano2010} RM formulae is the stellar line width in the absence of rotational broadening $\beta$.
We computed this as the quadrature sum of the widths of the instrumental line profile and macroturbulence based on the relation from \citet{Va05}, and placed a Gaussian prior with width 1~\kms on this parameter.
When fitting the KPF RVs, we noted that the instrumental uncertainties were likely overestimated, perhaps due to the early developmental stage of the KPF-DRP \citep[see e.g.,][]{Handley2024} --- the standard deviation of the residuals to the best-fit RM model were 1.2~\ms, compared with the median uncertainties of 2.0~\ms.
As such, we fitted for an additional RV jitter term in terms of $\sigma_J^2$ and added in quadrature to the instrumental uncertainties, allowing $\sigma_{J,\mathrm{KPF}}^2$ to be negative.

To constrain the remaining planetary orbital parameters that affect the RM velocity anomaly, as well as the overall slope in the RVs that arises from the Keplerian orbit of the star about the planet-star center-of-mass, we jointly fitted the TESS photometry, \textit{KeplerCam} photometry, and Magellan/PFS RVs previously published in \citetalias{Yee2023}.
The six year time baseline of the TESS data along with the contemporaneous \textit{KeplerCam} photometry placed a very strong constraint on the mid-transit time at the epoch of the KPF observations, which was particularly helpful given the lack of pre-transit baseline in the KPF data.
We modeled the photometric timeseries data using the \citet{MandelAgol02} transit models as implemented in \texttt{batman} \citep{Batman_Kreidberg15}, assuming a quadratic limb-darkening law.
For the \textit{KeplerCam} data, we simultaneously detrended the light curve against airmass while fitting the transit model.
We computed limb-darkening coefficients for TESS, \textit{KeplerCam}, and KPF using the exoCTK web calculator\footnote{\url{https://exoctk.stsci.edu/limb_darkening}} given the stellar properties from \citetalias{Yee2023}, and placed Gaussian priors centered at those values and with width 0.1 to account for modeling uncertainties and the uncertainties in stellar properties.

The RVs were modeled with a Keplerian orbit using \texttt{radvel} \citep{Radvel_Fulton18}, allowing for independent RV offsets between the KPF and PFS RVs.
We fixed the planet's eccentricity to zero, consistent with \citetalias{Yee2023}.
We also included priors on the stellar radius (used to compute \vsini) and mass (used to derive $a/\Rstar$ from the orbital period) based on a fit of the MIST bolometric correction tables \citep{MISTI_Choi2016} to broadband \textit{Gaia}, 2MASS and WISE photometry using the \texttt{EXOFASTv2} code \citep{ExoFASTv2_Eastman19}.
We placed an error floor of 4.2\% on \Rstar and added a systematic uncertainty of 5\% to the uncertainty in \Mstar, as recommended by \citet{Tayar2022} to capture realistic uncertainties in the absolute stellar radius scale and differences in stellar model grids.
Uninformative priors were used for the remaining fitting parameters.

\begin{figure*}
    \centering
    \includegraphics[width=0.9\textwidth]{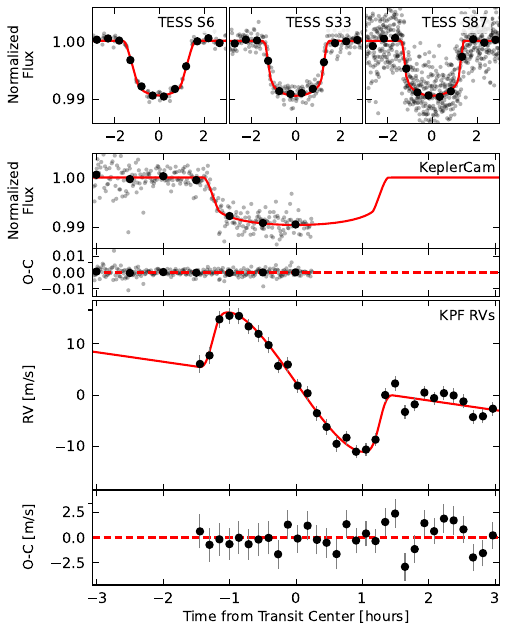}
    \caption{Results from joint modeling of photometric and RV observations. In all panels, the red line shows the best-fit model. The light gray dots show unbinned photometric data, while the black points in the top two rows show the same data binned to 1800s cadence. The TESS data have been phase-folded according to the best-fit transit ephemeris. For the RV data, unbinned data are shown as black points, and an arbitrary offset has been removed. The error bars for the KPF data have been reduced according to the best-fit jitter value.
    The data behind this figure are available in machine-readable form.}
    \label{fig:rm}
\end{figure*}

We show the transit photometry and KPF RVs in Figure \ref{fig:rm}, along with the best-fit model.
The KPF RVs showed a clear detection of the RM effect, and we measure a low sky-projected obliquity of $\lambda = \FitLambda{^\circ}$, and $\vsini = \FitVsini~\kms$.
Given nominal values for $\Prot = 23.5$~days and $\Rstar = 0.87\,\Rsun$, the expected equatorial rotation velocity for the star would be $v_\mathrm{eq} = 1.9\,\kms$, so the \vsini measurement from the RM effect suggests that $\sin{i_\star} \approx 1$.
Indeed, our MCMC analysis finds $\istar = \FitIstar{^\circ}$, which when combined with the measurement of $\lambda$, allows us to infer the true obliquity $\psi = \FitPsi{^\circ}$.
We note that because $\psi \geq |\lambda|$, any uncertainties in the angles $\lambda$, \istar, or \iorb will bias the median of the posterior probability distribution of $\psi$ upward from its true value especially when $\lambda$ is small.
Thus, we also report upper limits on $\psi$ (lower limits on $\cos{\psi}$) from our observations, finding $\psi < 19.2^\circ$ ($\cos{\psi} > 0.944$) with 68\% confidence and $\psi < 28.5^\circ$ ($\cos{\psi} > 0.879$) at 95\% confidence.
These results indicate that the orbit of TOI-2364\,b is likely well-aligned with the spin axis of its host star.

\begin{deluxetable*}{lcc} \label{tab:rm_results}
\tablecaption{Priors and Fit Results for RM Analysis}
\tablehead{\colhead{Parameter} & \colhead{Prior} & \colhead{Posterior}}
\startdata
$T_c$ (BJD$_\mathrm{TDB}$) & $\mathcal{U}(2460240.00,2460240.05)$ & $2460240.01995^{+0.00036}_{-0.00035}$ \\
$P_\mathrm{orb}$ (days) & $\mathcal{U}(4.0197,4.0198)$ & $4.0197478 \pm 0.0000014$ \\
$\lambda$ (deg) & $\mathcal{U}(-180,180)$ & $7^{+10}_{-11}$ \\
$\psi$ (deg) & Derived & $15.6^{+7.7}_{-7.3}$ \\
$\cos{i_\star}$ & $\mathcal{U}(-1,1)$ & $0.00 \pm 0.22$ \\
$P_\mathrm{rot}$ (days) & $\mathcal{N}(23.47,0.3)$ & $23.42 \pm 0.29$ \\
$v\sin{i_\star}$ (km/s) & Derived & $1.840^{+0.054}_{-0.060}$ \\
$b$ & $\mathcal{U}(0.0,1.0)$ & $0.13^{+0.11}_{-0.09}$ \\
$i_\mathrm{orb}$ (deg) & Derived & $89.38^{+0.42}_{-0.56}$ \\
$R_p/R_\star$ & $\mathcal{U}(0.08,0.10)$ & $0.09021^{+0.00081}_{-0.00080}$ \\
$R_\star$ ($R_\odot$) & $\mathcal{N}(0.873,0.037)$ & $0.868^{+0.020}_{-0.018}$ \\
$M_\star$ ($M_\odot$) & $\mathcal{N}(0.912,0.060)$ & $0.934^{+0.052}_{-0.051}$ \\
$\beta$ (km/s) & $\mathcal{N}(3.46,1.0)$ & $3.31^{+0.95}_{-0.84}$ \\
$K$ (m/s) & $\mathcal{U}(5.0,50.0)$ & $29.1^{+3.1}_{-3.0}$ \\
$\gamma_\mathrm{KPF}$ (km/s) & $\mathcal{U}(-1000,1000)$ & $2.71^{+0.46}_{-0.41}$ \\
$\sigma^2_{\mathrm{KPF}}$ (m/s) & $\mathcal{U}(-4,10)$ & $-1.88^{+0.67}_{-0.46}$ \\
$\gamma_\mathrm{PFS}$ (km/s) & $\mathcal{U}(-100,100)$ & $-14.5^{+3.1}_{-3.2}$ \\
$\sigma^2_{\mathrm{PFS}}$ (m/s) & $\mathcal{U}(0,400)$ & $46^{+84}_{-30}$ \\
$u_{1,\mathrm{KPF}}$ & $\mathcal{N}(0.58,0.1)$ & $0.556^{+0.085}_{-0.086}$ \\
$u_{2,\mathrm{KPF}}$ & $\mathcal{N}(0.13,0.1)$ & $0.114^{+0.093}_{-0.092}$ \\
$u_{1,\mathrm{TESS}}$ & $\mathcal{N}(0.454,0.1)$ & $0.395 \pm 0.059$ \\
$u_{2,\mathrm{TESS}}$ & $\mathcal{N}(0.172,0.1)$ & $0.133^{+0.088}_{-0.087}$ \\
$u_{1,i^{\prime}}$ & $\mathcal{N}(0.459,0.1)$ & $0.388^{+0.071}_{-0.072}$ \\
$u_{2,i^{\prime}}$ & $\mathcal{N}(0.17,0.1)$ & $0.139^{+0.091}_{-0.090}$ \\
$F_{0,\mathrm{TESS\,S6}}$ & $\mathcal{U}(0.99,1.01)$ & $1.000193 \pm 0.000082$ \\
$F_{0,\mathrm{TESS\,S33}}$ & $\mathcal{U}(0.99,1.01)$ & $1.000150 \pm 0.000074$ \\
$F_{0,\mathrm{TESS\,S87}}$ & $\mathcal{U}(0.99,1.01)$ & $1.00017 \pm 0.00011$ \\
$F_{0,\mathrm{Keplercam}}$ & $\mathcal{U}(0.99,1.01)$ & $0.99795 \pm 0.00012$ \\
$C_{0,\mathrm{Keplercam}}$ & $\mathcal{U}(-0.003,0.003)$ & $0.00030 \pm 0.00035$\enddata

\end{deluxetable*}

Given the high precision of our RV measurements, we considered the possibility that the in-transit velocity anomaly may have been affected by higher-order effects, such as center-to-limb variation in the convective blueshift and/or differential rotation.
We investigated models incorporating a linear dependence of the convective blueshift on the limb angle $\mu$ \citep{Shporer2011} as well as models including a latitude-dependent subplanetary velocity over the course of the transit.
We found that such models did not improve the fit to the KPF RVs and were therefore disfavored in a Bayesian sense given the additional free parameters.
Given the good alignment between the planet's orbit and the stellar rotation axis, the planet's shadow does not traverse a wide range of stellar latitudes and so it is unsurprising that the RM anomaly was insensitive to differential rotation in this case.
We also found no evidence for variation in the local line profiles occulted by the planet over the course of the transit, which may indicate center-to-limb variations in convective blueshift \citep{Rubenzahl2024a}.

Differential rotation may also affect our measurement of the true obliquity $\psi$ by breaking the assumption that $\Prot = 2\pi \Rstar / v_\mathrm{eq}$.
If the observed photometric modulation is due to spots confined to a particular active latitude $\theta_\mathrm{lat}$, then the measured rotation period would be $\Prot = 2\pi \Rstar / v_\mathrm{eq}(1 - \alpha \sin^2\theta_\mathrm{lat})$, where $\alpha$ is the rotational shear.
For a solar-like $\alpha = 0.2$ \citep[e.g.,][]{Balona2016} and spots at mid-latitudes $\theta_\mathrm{lat} = 30^\circ$, this would cause a $\approx 5\%$ increase in $v_\mathrm{eq}$ compared with the assumption of solid body rotation.
To test the sensitivity of our measurement to assumptions of differential rotation, we inflated the uncertainty on our measured \Prot to 10\%.
In this scenario, the constraint on $\psi$ is weakened to an upper limit of $\psi < 26^\circ$ and $\psi < 38^\circ$ at 68\% and 95\% confidence.
Still, we note that such a change does not alter the qualitative conclusion of a well-aligned orbit for TOI-2364\,b.

\section{Discussion and Conclusions}\label{sec:discussion}

In this work, we measured the rotation period of TOI-2364 using \textit{Tierras} Observatory data collected over a baseline of over 150~nights. We used a QP-GP model to estimate the rotation period and its uncertainty, finding $P_\mathrm{rot} =~$\prot~days. Fitting an RV dataset from KPF taken during a transit of TOI-2364~b, we measured the sky-projected stellar obliquity of $\lambda = \FitLambda{^\circ}$. A combination of the two dataset reveals that the planet's orbit is well-aligned with its host star spin axis, with $\psi = \FitPsi{^\circ}$.

TOI-2364\,b orbits a cool star below the Kraft break, where it has been suggested that efficient tidal dissipation allows planets to realign their host stars' spin axes with their orbits \citep{Schlaufman2010,Winn2010b}.
However, TOI-2364\,b has a low planetary mass and orbits relatively far from its star ($a/R_\star = 11.8\pm0.3$), such that it is not expected to raise significant tides.
Following \citet{Attia2023}, we computed a tidal realignment parameter $\tau \approx (5\pm1)\times10^{-16}$, below the threshold of $\tau\sim10^{-15}$ above which those authors suggested that tidal realignment becomes important.
Thus, the low obliquity of TOI-2364 is likely to be primordial rather than the result of tidal realignment.

\begin{figure*}
    \centering
    \includegraphics[width=\textwidth]{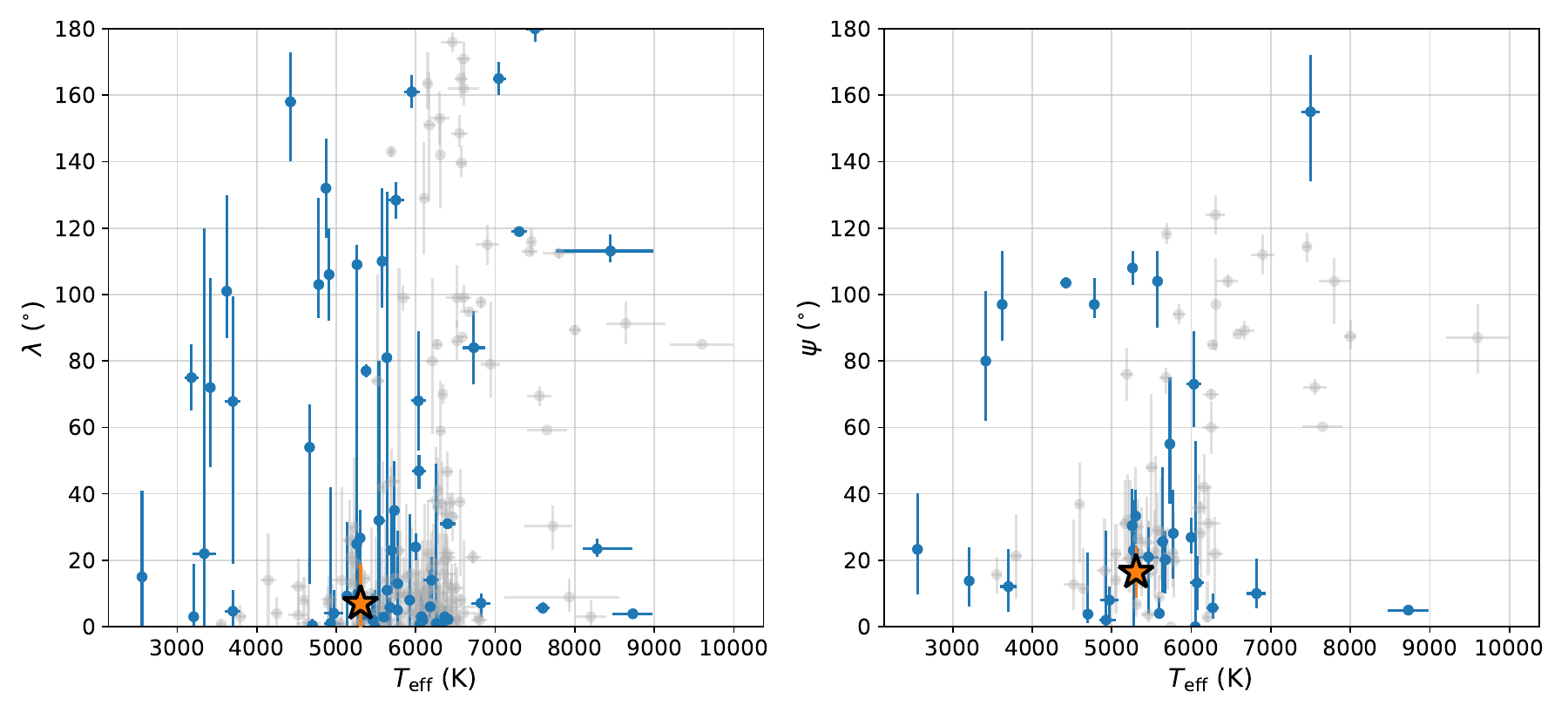}
    \caption{Measurements of the sky-projected obliquity $\lambda$ (left) and the true three-dimensional obliquity $\psi$ (right) for exoplanetary systems as a function of host star effective temperature. Planets with masses less than that of Saturn are colored in blue, while planets more massive than Saturn are shown in gray. Our measurements for TOI-2364 are indicated with an orange star. The possible bimodality of the orbits of planets less massive than Saturn around cool stars is more clearly revealed in the true obliquity distribution than in the sky-projected distribution.
    Measurements were compiled from \citet{Albrecht2022}, \citet{Knudstrup2024}, and the \texttt{TEPCat} catalog \citep{Southworth2011}, as of 28 Feb 2025.}
    \label{fig:obliquity_population}
\end{figure*}

In Figure~\ref{fig:obliquity_population} we show our measurements of $\lambda$ and $\psi$ for TOI-2364 in the context of the full population of obliquity measurements of exoplanetary systems as compiled from \citet{Albrecht2022}, \citet{Knudstrup2024}, and the \texttt{TEPCat} catalog \citep{Southworth2011}\footnote{\url{https://www.astro.keele.ac.uk/jkt/tepcat/obliquity.html}}. In this figure, planets with masses less than that of Saturn (of which TOI-2364 is one) are indicated in blue, while planets more massive than Saturn are shown in gray. As this figure reveals and as has been suggested in the literature, the ``preponderance of polar orbits'' \citep{Albrecht2021} is a phenomenon that may be particular to planets less massive than Saturn around cool stars \citep[with $T_\mathrm{eff} < 6500$~K;][]{Attia2023, Knudstrup2024, Espinoza-Retamal2024, Handley2024}.
\citet{Handley2024} suggested that planets with $M_p/M_\star\sim10^{-4}$ were more likely to exhibit this bimodality in planetary alignments; with $M_p/M_\star = 1.8\times10^{-4}$, TOI-2364\,b appears to be a member of the well-aligned subset of this population.

Figure \ref{fig:obliquity_population} also shows that the proposed bimodality in the distributions of these planets' orbits is more clearly revealed in the distribution of $\psi$ than in the distribution of $\lambda$. More measurements of $\psi$, particularly of sub-Saturn-mass planets, are needed to determine whether this apparent bimodality is genuine. 

In Section~\ref{subsec:prot}, we found that we maintained a night-to-night precision of \measuredSigNtoN~ppt over the entire \textit{Tierras} light curve, a precision that will enable rotation period measurements for other Sun-like stars, which have typical rotation periods of weeks and variability amplitudes of less than a percent. We are using \textit{Tierras} to conduct a search for rotation periods of stars with with potentially misaligned transiting planets, with the goal of increasing the population of systems for which the true obliquity is measured. This will determine whether there indeed exists a bimodal distribution of planet obliquities for short-period sub-Saturns around cool stars.

\begin{acknowledgments}
The \textit{Tierras} Observatory is supported by the National Science Foundation under Award No. AST-2308043. S.W.Y. and J.G.-M. gratefully acknowledge support from the Heising-Simons Foundation. J.G-M. acknowledges support from the Pappalardo family through the MIT Pappalardo Fellowship in Physics.
This work was supported by a NASA Keck PI Data Award, administered by the NASA Exoplanet Science Institute. Data presented herein were obtained at the W. M. Keck Observatory from telescope time allocated to the National Aeronautics and Space Administration through the agency's scientific partnership with the California Institute of Technology and the University of California. The Observatory was made possible by the generous financial support of the W. M. Keck Foundation. The authors wish to recognize and acknowledge the very significant cultural role and reverence that the summit of Maunakea has always had within the indigenous Hawaiian community. We are most fortunate to have the opportunity to conduct observations from this mountain.
Some of the data presented in this paper were obtained from the Mikulski Archive for Space Telescopes (MAST) at the Space Telescope Science Institute. The specific observations analyzed can be accessed via MAST \citep{MAST2021a, MAST2021b}.
\end{acknowledgments}

\vspace{5mm}
\facilities{All-Sky Automated Survey for Supernovae (ASAS-SN) \citep{Shappee2014}, 
Transiting Exoplanet Survey Satellite (TESS) \citep{Ri15},
Zwicky Transient Facility (ZTF) \citep{Bellm2018},
\textit{Tierras} Observatory \citep{GaMe20}, 
Keck:I (KPF) \citep{Gibson2024}
 }

\software{\texttt{TEPCat} \citep{Southworth2011},
\texttt{emcee} \citep{Fo13}, 
\texttt{photutils} \citep{Bradley2023},
\texttt{rmfit} \citep{rmfit_Stefansson2022},
\texttt{radvel} \citep{Radvel_Fulton18},
\texttt{batman} \citep{Batman_Kreidberg15},
\texttt{EXOFASTv2} \citep{ExoFASTv2_Eastman19}
}

\vspace{5mm}


\bibliography{bib_file}{}

\begin{thebibliography}{}
\expandafter\ifx\csname natexlab\endcsname\relax\def\natexlab#1{#1}\fi
\providecommand{\url}[1]{\href{#1}{#1}}
\providecommand{\dodoi}[1]{doi:~\href{http://doi.org/#1}{\nolinkurl{#1}}}
\providecommand{\doeprint}[1]{\href{http://ascl.net/#1}{\nolinkurl{http://ascl.net/#1}}}
\providecommand{\doarXiv}[1]{\href{https://arxiv.org/abs/#1}{\nolinkurl{https://arxiv.org/abs/#1}}}

\bibitem[{{Albrecht} {et~al.}(2022){Albrecht}, {Dawson}, \& {Winn}}]{Albrecht2022}
{Albrecht}, S.~H., {Dawson}, R.~I., \& {Winn}, J.~N. 2022, \pasp, 134, 082001, \dodoi{10.1088/1538-3873/ac6c09}

\bibitem[{{Albrecht} {et~al.}(2021){Albrecht}, {Marcussen}, {Winn}, {Dawson}, \& {Knudstrup}}]{Albrecht2021}
{Albrecht}, S.~H., {Marcussen}, M.~L., {Winn}, J.~N., {Dawson}, R.~I., \& {Knudstrup}, E. 2021, \apjl, 916, L1, \dodoi{10.3847/2041-8213/ac0f03}

\bibitem[{{Angus} {et~al.}(2018){Angus}, {Morton}, {Aigrain}, {Foreman-Mackey}, \& {Rajpaul}}]{Angus2018}
{Angus}, R., {Morton}, T., {Aigrain}, S., {Foreman-Mackey}, D., \& {Rajpaul}, V. 2018, \mnras, 474, 2094, \dodoi{10.1093/mnras/stx2109}

\bibitem[{{Astropy Collaboration} {et~al.}(2013){Astropy Collaboration}, {Robitaille}, {Tollerud}, {Greenfield}, {Droettboom}, {Bray}, {Aldcroft}, {Davis}, {Ginsburg}, {Price-Whelan}, {Kerzendorf}, {Conley}, {Crighton}, {Barbary}, {Muna}, {Ferguson}, {Grollier}, {Parikh}, {Nair}, {Unther}, {Deil}, {Woillez}, {Conseil}, {Kramer}, {Turner}, {Singer}, {Fox}, {Weaver}, {Zabalza}, {Edwards}, {Azalee Bostroem}, {Burke}, {Casey}, {Crawford}, {Dencheva}, {Ely}, {Jenness}, {Labrie}, {Lim}, {Pierfederici}, {Pontzen}, {Ptak}, {Refsdal}, {Servillat}, \& {Streicher}}]{astropy:2013}
{Astropy Collaboration}, {Robitaille}, T.~P., {Tollerud}, E.~J., {et~al.} 2013, \aap, 558, A33, \dodoi{10.1051/0004-6361/201322068}

\bibitem[{{Astropy Collaboration} {et~al.}(2018){Astropy Collaboration}, {Price-Whelan}, {Sip{\H{o}}cz}, {G{\"u}nther}, {Lim}, {Crawford}, {Conseil}, {Shupe}, {Craig}, {Dencheva}, {Ginsburg}, {Vand erPlas}, {Bradley}, {P{\'e}rez-Su{\'a}rez}, {de Val-Borro}, {Aldcroft}, {Cruz}, {Robitaille}, {Tollerud}, {Ardelean}, {Babej}, {Bach}, {Bachetti}, {Bakanov}, {Bamford}, {Barentsen}, {Barmby}, {Baumbach}, {Berry}, {Biscani}, {Boquien}, {Bostroem}, {Bouma}, {Brammer}, {Bray}, {Breytenbach}, {Buddelmeijer}, {Burke}, {Calderone}, {Cano Rodr{\'\i}guez}, {Cara}, {Cardoso}, {Cheedella}, {Copin}, {Corrales}, {Crichton}, {D'Avella}, {Deil}, {Depagne}, {Dietrich}, {Donath}, {Droettboom}, {Earl}, {Erben}, {Fabbro}, {Ferreira}, {Finethy}, {Fox}, {Garrison}, {Gibbons}, {Goldstein}, {Gommers}, {Greco}, {Greenfield}, {Groener}, {Grollier}, {Hagen}, {Hirst}, {Homeier}, {Horton}, {Hosseinzadeh}, {Hu}, {Hunkeler}, {Ivezi{\'c}}, {Jain}, {Jenness}, {Kanarek}, {Kendrew}, {Kern}, {Kerzendorf}, {Khvalko}, {King}, {Kirkby}, {Kulkarni},
  {Kumar}, {Lee}, {Lenz}, {Littlefair}, {Ma}, {Macleod}, {Mastropietro}, {McCully}, {Montagnac}, {Morris}, {Mueller}, {Mumford}, {Muna}, {Murphy}, {Nelson}, {Nguyen}, {Ninan}, {N{\"o}the}, {Ogaz}, {Oh}, {Parejko}, {Parley}, {Pascual}, {Patil}, {Patil}, {Plunkett}, {Prochaska}, {Rastogi}, {Reddy Janga}, {Sabater}, {Sakurikar}, {Seifert}, {Sherbert}, {Sherwood-Taylor}, {Shih}, {Sick}, {Silbiger}, {Singanamalla}, {Singer}, {Sladen}, {Sooley}, {Sornarajah}, {Streicher}, {Teuben}, {Thomas}, {Tremblay}, {Turner}, {Terr{\'o}n}, {van Kerkwijk}, {de la Vega}, {Watkins}, {Weaver}, {Whitmore}, {Woillez}, {Zabalza}, \& {Astropy Contributors}}]{astropy:2018}
{Astropy Collaboration}, {Price-Whelan}, A.~M., {Sip{\H{o}}cz}, B.~M., {et~al.} 2018, \aj, 156, 123, \dodoi{10.3847/1538-3881/aabc4f}

\bibitem[{{Astropy Collaboration} {et~al.}(2022){Astropy Collaboration}, {Price-Whelan}, {Lim}, {Earl}, {Starkman}, {Bradley}, {Shupe}, {Patil}, {Corrales}, {Brasseur}, {N{"o}the}, {Donath}, {Tollerud}, {Morris}, {Ginsburg}, {Vaher}, {Weaver}, {Tocknell}, {Jamieson}, {van Kerkwijk}, {Robitaille}, {Merry}, {Bachetti}, {G{"u}nther}, {Aldcroft}, {Alvarado-Montes}, {Archibald}, {B{'o}di}, {Bapat}, {Barentsen}, {Baz{'a}n}, {Biswas}, {Boquien}, {Burke}, {Cara}, {Cara}, {Conroy}, {Conseil}, {Craig}, {Cross}, {Cruz}, {D'Eugenio}, {Dencheva}, {Devillepoix}, {Dietrich}, {Eigenbrot}, {Erben}, {Ferreira}, {Foreman-Mackey}, {Fox}, {Freij}, {Garg}, {Geda}, {Glattly}, {Gondhalekar}, {Gordon}, {Grant}, {Greenfield}, {Groener}, {Guest}, {Gurovich}, {Handberg}, {Hart}, {Hatfield-Dodds}, {Homeier}, {Hosseinzadeh}, {Jenness}, {Jones}, {Joseph}, {Kalmbach}, {Karamehmetoglu}, {Ka{l}uszy{'n}ski}, {Kelley}, {Kern}, {Kerzendorf}, {Koch}, {Kulumani}, {Lee}, {Ly}, {Ma}, {MacBride}, {Maljaars}, {Muna}, {Murphy}, {Norman}, {O'Steen},
  {Oman}, {Pacifici}, {Pascual}, {Pascual-Granado}, {Patil}, {Perren}, {Pickering}, {Rastogi}, {Roulston}, {Ryan}, {Rykoff}, {Sabater}, {Sakurikar}, {Salgado}, {Sanghi}, {Saunders}, {Savchenko}, {Schwardt}, {Seifert-Eckert}, {Shih}, {Jain}, {Shukla}, {Sick}, {Simpson}, {Singanamalla}, {Singer}, {Singhal}, {Sinha}, {Sip{H{o}}cz}, {Spitler}, {Stansby}, {Streicher}, {{{S}}umak}, {Swinbank}, {Taranu}, {Tewary}, {Tremblay}, {Val-Borro}, {Van Kooten}, {Vasovi{'c}}, {Verma}, {de Miranda Cardoso}, {Williams}, {Wilson}, {Winkel}, {Wood-Vasey}, {Xue}, {Yoachim}, {Zhang}, {Zonca}, \& {Astropy Project Contributors}}]{astropy:2022}
{Astropy Collaboration}, {Price-Whelan}, A.~M., {Lim}, P.~L., {et~al.} 2022, \apj, 935, 167, \dodoi{10.3847/1538-4357/ac7c74}

\bibitem[{Attia {et~al.}(2023)Attia, Bourrier, Delisle, \& Eggenberger}]{Attia2023}
Attia, O., Bourrier, V., Delisle, J.-B., \& Eggenberger, P. 2023, {{DREAM II}}. {{The}} Spin-Orbit Angle Distribution of Close-in Exoplanets under the Lens of Tides,  arXiv, \dodoi{10.48550/arXiv.2305.00829}

\bibitem[{{Balona} \& {Abedigamba}(2016)}]{Balona2016}
{Balona}, L.~A., \& {Abedigamba}, O.~P. 2016, \mnras, 461, 497, \dodoi{10.1093/mnras/stw1443}

\bibitem[{{Baluev}(2008)}]{Baluev2008}
{Baluev}, R.~V. 2008, \mnras, 385, 1279, \dodoi{10.1111/j.1365-2966.2008.12689.x}

\bibitem[{{Baranne} {et~al.}(1996){Baranne}, {Queloz}, {Mayor}, {Adrianzyk}, {Knispel}, {Kohler}, {Lacroix}, {Meunier}, {Rimbaud}, \& {Vin}}]{Ba96}
{Baranne}, A., {Queloz}, D., {Mayor}, M., {et~al.} 1996, \aaps, 119, 373

\bibitem[{Bellm {et~al.}(2018)Bellm, Kulkarni, Graham, Dekany, Smith, Riddle, Masci, Helou, Prince, Adams, Barbarino, Barlow, Bauer, Beck, Belicki, Biswas, Blagorodnova, Bodewits, Bolin, Brinnel, Brooke, Bue, Bulla, Burruss, Cenko, Chang, Connolly, Coughlin, Cromer, Cunningham, De, Delacroix, Desai, Duev, Eadie, Farnham, Feeney, Feindt, Flynn, Franckowiak, Frederick, Fremling, Gal-Yam, Gezari, Giomi, Goldstein, Golkhou, Goobar, Groom, Hacopians, Hale, Henning, Ho, Hover, Howell, Hung, Huppenkothen, Imel, Ip, Ivezić, Jackson, Jones, Juric, Kasliwal, Kaspi, Kaye, Kelley, Kowalski, Kramer, Kupfer, Landry, Laher, Lee, Lin, Lin, Lunnan, Giomi, Mahabal, Mao, Miller, Monkewitz, Murphy, Ngeow, Nordin, Nugent, Ofek, Patterson, Penprase, Porter, Rauch, Rebbapragada, Reiley, Rigault, Rodriguez, Roestel, Rusholme, Santen, Schulze, Shupe, Singer, Soumagnac, Stein, Surace, Sollerman, Szkody, Taddia, Terek, Van~Sistine, van Velzen, Vestrand, Walters, Ward, Ye, Yu, Yan, \& Zolkower}]{Bellm2018}
Bellm, E.~C., Kulkarni, S.~R., Graham, M.~J., {et~al.} 2018, Publications of the Astronomical Society of the Pacific, 131, 018002, \dodoi{10.1088/1538-3873/aaecbe}

\bibitem[{{Borucki} {et~al.}(2010){Borucki}, {Koch}, {Basri}, {Batalha}, {Brown}, {Caldwell}, {Caldwell}, {Christensen-Dalsgaard}, {Cochran}, {DeVore}, {Dunham}, {Dupree}, {Gautier}, {Geary}, {Gilliland}, {Gould}, {Howell}, {Jenkins}, {Kondo}, {Latham}, {Marcy}, {Meibom}, {Kjeldsen}, {Lissauer}, {Monet}, {Morrison}, {Sasselov}, {Tarter}, {Boss}, {Brownlee}, {Owen}, {Buzasi}, {Charbonneau}, {Doyle}, {Fortney}, {Ford}, {Holman}, {Seager}, {Steffen}, {Welsh}, {Rowe}, {Anderson}, {Buchhave}, {Ciardi}, {Walkowicz}, {Sherry}, {Horch}, {Isaacson}, {Everett}, {Fischer}, {Torres}, {Johnson}, {Endl}, {MacQueen}, {Bryson}, {Dotson}, {Haas}, {Kolodziejczak}, {Van Cleve}, {Chandrasekaran}, {Twicken}, {Quintana}, {Clarke}, {Allen}, {Li}, {Wu}, {Tenenbaum}, {Verner}, {Bruhweiler}, {Barnes}, \& {Prsa}}]{Bo10a}
{Borucki}, W.~J., {Koch}, D., {Basri}, G., {et~al.} 2010, Science, 327, 977, \dodoi{10.1126/science.1185402}

\bibitem[{Bradley {et~al.}(2023)Bradley, Sip{\H o}cz, Robitaille, Tollerud, Vin{\'{\i}}cius, Deil, Barbary, Wilson, Busko, Donath, G{\"u}nther, Cara, Lim, Me{\ss}linger, Conseil, Bostroem, Droettboom, Bray, Bratholm, Barentsen, Craig, Rathi, Pascual, Perren, Georgiev, de~Val-Borro, Kerzendorf, Bach, Quint, \& Souchereau}]{Bradley2023}
Bradley, L., Sip{\H o}cz, B., Robitaille, T., {et~al.} 2023, astropy/photutils: 1.8.0, 1.8.0,  Zenodo, \dodoi{10.5281/zenodo.7946442}

\bibitem[{Choi {et~al.}(2016)Choi, Dotter, Conroy, Cantiello, Paxton, \& Johnson}]{MISTI_Choi2016}
Choi, J., Dotter, A., Conroy, C., {et~al.} 2016, The Astrophysical Journal, 823, 102, \dodoi{10.3847/0004-637X/823/2/102}

\bibitem[{{Claytor} {et~al.}(2024){Claytor}, {van Saders}, {Cao}, {Pinsonneault}, {Teske}, \& {Beaton}}]{Claytor2024}
{Claytor}, Z.~R., {van Saders}, J.~L., {Cao}, L., {et~al.} 2024, \apj, 962, 47, \dodoi{10.3847/1538-4357/ad159a}

\bibitem[{{Collins} {et~al.}(2017){Collins}, {Kielkopf}, {Stassun}, \& {Hessman}}]{Co17}
{Collins}, K.~A., {Kielkopf}, J.~F., {Stassun}, K.~G., \& {Hessman}, F.~V. 2017, \aj, 153, 77, \dodoi{10.3847/1538-3881/153/2/77}

\bibitem[{Dong \& {Foreman-Mackey}(2023)}]{Dong2023}
Dong, J., \& {Foreman-Mackey}, D. 2023, The Astronomical Journal, 166, 112, \dodoi{10.3847/1538-3881/ace105}

\bibitem[{{Eastman} {et~al.}(2019){Eastman}, {Rodriguez}, {Agol}, {Stassun}, {Beatty}, {Vanderburg}, {Gaudi}, {Collins}, \& {Luger}}]{ExoFASTv2_Eastman19}
{Eastman}, J.~D., {Rodriguez}, J.~E., {Agol}, E., {et~al.} 2019, arXiv e-prints, arXiv:1907.09480.
\newblock \doarXiv{1907.09480}

\bibitem[{{Espinoza-Retamal} {et~al.}(2024){Espinoza-Retamal}, Stef{\'a}nsson, Petrovich, Brahm, Jord{\'a}n, Sedaghati, Lucero, Tala~Pinto, Mu{\~n}oz, Boyle, Leiva, \& Suc}]{Espinoza-Retamal2024}
{Espinoza-Retamal}, J.~I., Stef{\'a}nsson, G., Petrovich, C., {et~al.} 2024, The Astronomical Journal, 168, \dodoi{10.3847/1538-3881/ad70b8}

\bibitem[{{Foreman-Mackey} {et~al.}(2017){Foreman-Mackey}, {Agol}, {Ambikasaran}, \& {Angus}}]{ForemanMackey2017}
{Foreman-Mackey}, D., {Agol}, E., {Ambikasaran}, S., \& {Angus}, R. 2017, \aj, 154, 220, \dodoi{10.3847/1538-3881/aa9332}

\bibitem[{{Foreman-Mackey} {et~al.}(2013){Foreman-Mackey}, {Hogg}, {Lang}, \& {Goodman}}]{Fo13}
{Foreman-Mackey}, D., {Hogg}, D.~W., {Lang}, D., \& {Goodman}, J. 2013, \pasp, 125, 306, \dodoi{10.1086/670067}

\bibitem[{{Fulton} {et~al.}(2018){Fulton}, {Petigura}, {Blunt}, \& {Sinukoff}}]{Radvel_Fulton18}
{Fulton}, B.~J., {Petigura}, E.~A., {Blunt}, S., \& {Sinukoff}, E. 2018, \pasp, 130, 044504, \dodoi{10.1088/1538-3873/aaaaa8}

\bibitem[{{García-Mejía} {et~al.}(2020){García-Mejía}, {Charbonneau}, {Fabricant}, {Irwin}, {Fata}, {Zajac}, \& {Doherty}}]{GaMe20}
{García-Mejía}, J., {Charbonneau}, D., {Fabricant}, D., {et~al.} 2020, in Society of Photo-Optical Instrumentation Engineers (SPIE) Conference Series, Vol. 11445, Ground-based and Airborne Telescopes VIII, ed. H.~K. {Marshall}, J.~{Spyromilio}, \& T.~{Usuda}, 114457R, \dodoi{10.1117/12.2561467}

\bibitem[{{Gibson} {et~al.}(2020){Gibson}, {Howard}, {Rider}, {Roy}, {Edelstein}, {Kassis}, {Grillo}, {Halverson}, {Sirk}, {Smith}, {Allen}, {Baker}, {Beichman}, {Berriman}, {Brown}, {Casey}, {Chin}, {Coutts}, {Cowley}, {Deich}, {Feger}, {Fulton}, {Gers}, {Gurevich}, {Ishikawa}, {James}, {Jelinsky}, {Kaye}, {Lanclos}, {Li}, {Lilley}, {McCarney}, {Miller}, {Milner}, {O'Hanlon}, {Pember}, {Raffanti}, {Rockosi}, {Rubenzahl}, {Rumph}, {Sandford}, {Savage}, {Schwab}, {Seifahrt}, {Shaum}, {Smith}, {Stuermer}, {Thorne}, {Vandenberg}, {Von Boeckmann}, {Wang}, {Wang}, {Weisfeiler}, {Wilcox}, {Wishnow}, {Wizinowich}, {Wold}, \& {Wolfenberger}}]{Gibson2020}
{Gibson}, S.~R., {Howard}, A.~W., {Rider}, K., {et~al.} 2020, in Society of Photo-Optical Instrumentation Engineers (SPIE) Conference Series, Vol. 11447, Society of Photo-Optical Instrumentation Engineers (SPIE) Conference Series, 1144742, \dodoi{10.1117/12.2561783}

\bibitem[{{Gibson} {et~al.}(2024){Gibson}, {Howard}, {Rider}, {Halverson}, {Roy}, {Baker}, {Edelstein}, {Smith}, {Fulton}, {Walawender}, {Brodheim}, {Brown}, {Chan}, {Dai}, {Deich}, {Gottschalk}, {Grillo}, {Hale}, {Hill}, {Holden}, {Householder}, {Isaacson}, {Ishikawa}, {Jelinsky}, {Kassis}, {Kaye}, {Laher}, {Lanclos}, {Lee}, {Lilley}, {McCarney}, {Miller}, {Payne}, {Petigura}, {Poppett}, {Raffanti}, {Rubenzahl}, {Sandford}, {Schwab}, {Shaum}, {Sirk}, {Smith}, {Thorne}, {Valliant}, {Vandenberg}, {Wang}, {Wishnow}, {Wold}, {Yeh}, {Baca}, {Beichman}, {Berriman}, {Brown}, {Casey}, {Chin}, {Chong}, {Cowley}, {Devenot}, {Elwir}, {Finstad}, {Fraysse}, {James}, {Jhoti}, {Killian}, {Levine}, {Li}, {Marin}, {Milner}, {Nance}, {O'Hanlon}, {Orr}, {Ortiz-Soto}, {Payne}, {Pember}, {Raskin}, {Savage}, {Seifahrt}, {Smith}, {Storesund}, {St{\"u}rmer}, {Suominen}, {Tehero}, {Von Boeckmann}, {Wages}, {Weisfeiler}, {Wilcox}, {Wizinowich}, \& {Wolfenberger}}]{Gibson2024}
{Gibson}, S.~R., {Howard}, A.~W., {Rider}, K., {et~al.} 2024, in Society of Photo-Optical Instrumentation Engineers (SPIE) Conference Series, Vol. 13096, Ground-based and Airborne Instrumentation for Astronomy X, ed. J.~J. {Bryant}, K.~{Motohara}, \& J.~R.~D. {Vernet}, 1309609, \dodoi{10.1117/12.3017841}

\bibitem[{{Gordon} {et~al.}(2021){Gordon}, {Davenport}, {Angus}, {Foreman-Mackey}, {Agol}, {Covey}, {Ag{\"u}eros}, \& {Kipping}}]{Gordon2021}
{Gordon}, T.~A., {Davenport}, J. R.~A., {Angus}, R., {et~al.} 2021, \apj, 913, 70, \dodoi{10.3847/1538-4357/abf63e}

\bibitem[{Handley {et~al.}(2024)Handley, Howard, Rubenzahl, Dai, Tyler, Lee, Giacalone, Isaacson, Householder, Halverson, Roy, \& Walawender}]{Handley2024}
Handley, L.~B., Howard, A.~W., Rubenzahl, R.~A., {et~al.} 2024, An {{Obliquity Measurement}} of the {{Hot Neptune TOI-1694b}}, \dodoi{10.48550/arXiv.2412.07950}

\bibitem[{{H{\'e}brard} {et~al.}(2008){H{\'e}brard}, {Bouchy}, {Pont}, {Loeillet}, {Rabus}, {Bonfils}, {Moutou}, {Boisse}, {Delfosse}, {Desort}, {Eggenberger}, {Ehrenreich}, {Forveille}, {Lagrange}, {Lovis}, {Mayor}, {Pepe}, {Perrier}, {Queloz}, {Santos}, {S{\'e}gransan}, {Udry}, \& {Vidal-Madjar}}]{Hebrard2008}
{H{\'e}brard}, G., {Bouchy}, F., {Pont}, F., {et~al.} 2008, \aap, 488, 763, \dodoi{10.1051/0004-6361:200810056}

\bibitem[{Hirano {et~al.}(2010)Hirano, Suto, Taruya, Narita, Sato, Johnson, \& Winn}]{Hirano2010}
Hirano, T., Suto, Y., Taruya, A., {et~al.} 2010, The Astrophysical Journal, 709, 458, \dodoi{10.1088/0004-637X/709/1/458}

\bibitem[{{Holcomb} {et~al.}(2022){Holcomb}, {Robertson}, {Hartigan}, {Oelkers}, \& {Robinson}}]{Holcomb2022}
{Holcomb}, R.~J., {Robertson}, P., {Hartigan}, P., {Oelkers}, R.~J., \& {Robinson}, C. 2022, \apj, 936, 138, \dodoi{10.3847/1538-4357/ac8990}

\bibitem[{{Howell} {et~al.}(2014){Howell}, {Sobeck}, {Haas}, {Still}, {Barclay}, {Mullally}, {Troeltzsch}, {Aigrain}, {Bryson}, {Caldwell}, {Chaplin}, {Cochran}, {Huber}, {Marcy}, {Miglio}, {Najita}, {Smith}, {Twicken}, \& {Fortney}}]{Ho14}
{Howell}, S.~B., {Sobeck}, C., {Haas}, M., {et~al.} 2014, \pasp, 126, 398, \dodoi{10.1086/676406}

\bibitem[{{Jenkins} {et~al.}(2016){Jenkins}, {Twicken}, {McCauliff}, {Campbell}, {Sanderfer}, {Lung}, {Mansouri-Samani}, {Girouard}, {Tenenbaum}, {Klaus}, {Smith}, {Caldwell}, {Chacon}, {Henze}, {Heiges}, {Latham}, {Morgan}, {Swade}, {Rinehart}, \& {Vanderspek}}]{Jenkins2016}
{Jenkins}, J.~M., {Twicken}, J.~D., {McCauliff}, S., {et~al.} 2016, in Society of Photo-Optical Instrumentation Engineers (SPIE) Conference Series, Vol. 9913, Software and Cyberinfrastructure for Astronomy IV, ed. G.~{Chiozzi} \& J.~C. {Guzman}, 99133E, \dodoi{10.1117/12.2233418}

\bibitem[{Knudstrup {et~al.}(2024)Knudstrup, Albrecht, Winn, Gandolfi, Zanazzi, Persson, Fridlund, Marcussen, Chontos, Keniger, Eisner, Bieryla, Isaacson, Howard, Hirsch, Murgas, Narita, Palle, Kawai, \& Baker}]{Knudstrup2024}
Knudstrup, E., Albrecht, S.~H., Winn, J.~N., {et~al.} 2024, Obliquities of {{Exoplanet Host Stars}}: 19 {{New}} and {{Updated Measurements}}, and {{Trends}} in the {{Sample}} of 205 {{Measurements}},  arXiv.
\newblock \doarXiv{2408.09793}

\bibitem[{{Kochanek} {et~al.}(2017){Kochanek}, {Shappee}, {Stanek}, {Holoien}, {Thompson}, {Prieto}, {Dong}, {Shields}, {Will}, {Britt}, {Perzanowski}, \& {Pojma{\'n}ski}}]{Kochanek2017}
{Kochanek}, C.~S., {Shappee}, B.~J., {Stanek}, K.~Z., {et~al.} 2017, \pasp, 129, 104502, \dodoi{10.1088/1538-3873/aa80d9}

\bibitem[{{Kreidberg}(2015)}]{Batman_Kreidberg15}
{Kreidberg}, L. 2015, ArXiv e-prints.
\newblock \doarXiv{1507.08285}

\bibitem[{{Lockwood} {et~al.}(2007){Lockwood}, {Skiff}, {Henry}, {Henry}, {Radick}, {Baliunas}, {Donahue}, \& {Soon}}]{Lockwood2007}
{Lockwood}, G.~W., {Skiff}, B.~A., {Henry}, G.~W., {et~al.} 2007, \apjs, 171, 260, \dodoi{10.1086/516752}

\bibitem[{{Mandel} \& {Agol}(2002)}]{MandelAgol02}
{Mandel}, K., \& {Agol}, E. 2002, \apjl, 580, L171, \dodoi{10.1086/345520}

\bibitem[{{MAST Team}(2021{\natexlab{a}})}]{MAST2021a}
{MAST Team}. 2021{\natexlab{a}}, TESS ``Fast" Light Curves - All Sectors,  STScI/MAST, \dodoi{10.17909/T9-ST5G-3177}

\bibitem[{{MAST Team}(2021{\natexlab{b}})}]{MAST2021b}
---. 2021{\natexlab{b}}, TESS Light Curves - All Sectors,  STScI/MAST, \dodoi{10.17909/T9-NMC8-F686}

\bibitem[{Masuda {et~al.}(2022)Masuda, Petigura, \& Hall}]{Masuda2022a}
Masuda, K., Petigura, E.~A., \& Hall, O.~J. 2022, Monthly Notices of the Royal Astronomical Society, 510, 5623, \dodoi{10.1093/mnras/stab3650}

\bibitem[{Masuda \& Winn(2020)}]{Masuda2020}
Masuda, K., \& Winn, J.~N. 2020, The Astronomical Journal, 159, 81, \dodoi{10.3847/1538-3881/ab65be}

\bibitem[{{McQuillan} {et~al.}(2014){McQuillan}, {Mazeh}, \& {Aigrain}}]{McQuillan2014}
{McQuillan}, A., {Mazeh}, T., \& {Aigrain}, S. 2014, \apjs, 211, 24, \dodoi{10.1088/0067-0049/211/2/24}

\bibitem[{{Nicholson} \& {Aigrain}(2022)}]{Nicholson2022}
{Nicholson}, B.~A., \& {Aigrain}, S. 2022, \mnras, 515, 5251, \dodoi{10.1093/mnras/stac2097}

\bibitem[{{Nielsen} {et~al.}(2013){Nielsen}, {Gizon}, {Schunker}, \& {Karoff}}]{Nielsen2013}
{Nielsen}, M.~B., {Gizon}, L., {Schunker}, H., \& {Karoff}, C. 2013, \aap, 557, L10, \dodoi{10.1051/0004-6361/201321912}

\bibitem[{Pepe {et~al.}(2002)Pepe, Mayor, Galland, Naef, Queloz, Santos, Udry, \& Burnet}]{Pepe2002}
Pepe, F., Mayor, M., Galland, F., {et~al.} 2002, Astronomy \& Astrophysics, 388, 632, \dodoi{10.1051/0004-6361:20020433}

\bibitem[{{Petigura}(2015)}]{SpecMatchSynth_Petigura2015}
{Petigura}, E.~A. 2015, PhD thesis, University of California, Berkeley, United States

\bibitem[{Petrovich {et~al.}(2020)Petrovich, Mu{\~n}oz, Kratter, \& Malhotra}]{Petrovich2020}
Petrovich, C., Mu{\~n}oz, D.~J., Kratter, K.~M., \& Malhotra, R. 2020, The Astrophysical Journal, 902, \dodoi{10.3847/2041-8213/abb952}

\bibitem[{{Reinhold} \& {Hekker}(2020)}]{Reinhold2020}
{Reinhold}, T., \& {Hekker}, S. 2020, \aap, 635, A43, \dodoi{10.1051/0004-6361/201936887}

\bibitem[{{Ricker} {et~al.}(2015){Ricker}, {Winn}, {Vanderspek}, {Latham}, {Bakos}, {Bean}, {Berta-Thompson}, {Brown}, {Buchhave}, {Butler}, {Butler}, {Chaplin}, {Charbonneau}, {Christensen-Dalsgaard}, {Clampin}, {Deming}, {Doty}, {De Lee}, {Dressing}, {Dunham}, {Endl}, {Fressin}, {Ge}, {Henning}, {Holman}, {Howard}, {Ida}, {Jenkins}, {Jernigan}, {Johnson}, {Kaltenegger}, {Kawai}, {Kjeldsen}, {Laughlin}, {Levine}, {Lin}, {Lissauer}, {MacQueen}, {Marcy}, {McCullough}, {Morton}, {Narita}, {Paegert}, {Palle}, {Pepe}, {Pepper}, {Quirrenbach}, {Rinehart}, {Sasselov}, {Sato}, {Seager}, {Sozzetti}, {Stassun}, {Sullivan}, {Szentgyorgyi}, {Torres}, {Udry}, \& {Villasenor}}]{Ri15}
{Ricker}, G.~R., {Winn}, J.~N., {Vanderspek}, R., {et~al.} 2015, Journal of Astronomical Telescopes, Instruments, and Systems, 1, 014003, \dodoi{10.1117/1.JATIS.1.1.014003}

\bibitem[{Rubenzahl {et~al.}(2024)Rubenzahl, Dai, Halverson, Howard, Householder, Fulton, Behmard, Gibson, Roy, Shaum, Isaacson, Brodheim, Deich, Hill, Holden, Laher, Lanclos, Payne, Petigura, Schwab, Smith, Stef{\'a}nsson, Walawender, Wang, Weiss, Winn, \& Wishnow}]{Rubenzahl2024a}
Rubenzahl, R.~A., Dai, F., Halverson, S., {et~al.} 2024, The Astronomical Journal, 168, 188, \dodoi{10.3847/1538-3881/ad70b5}

\bibitem[{Schlaufman(2010)}]{Schlaufman2010}
Schlaufman, K.~C. 2010, The Astrophysical Journal, 719, 602, \dodoi{10.1088/0004-637X/719/1/602}

\bibitem[{Shallue \& Vanderburg(2018)}]{Keplerspline_Shallue2018}
Shallue, C.~J., \& Vanderburg, A. 2018, The Astronomical Journal, 155, 94, \dodoi{10.3847/1538-3881/aa9e09}

\bibitem[{{Shappee} {et~al.}(2014){Shappee}, {Prieto}, {Stanek}, {Kochanek}, {Holoien}, {Jencson}, {Basu}, {Beacom}, {Szczygiel}, {Pojmanski}, {Brimacombe}, {Dubberley}, {Elphick}, {Foale}, {Hawkins}, {Mullins}, {Rosing}, {Ross}, \& {Walker}}]{Shappee2014}
{Shappee}, B., {Prieto}, J., {Stanek}, K.~Z., {et~al.} 2014, in American Astronomical Society Meeting Abstracts, Vol. 223, American Astronomical Society Meeting Abstracts \#223, 236.03

\bibitem[{Shporer \& Brown(2011)}]{Shporer2011}
Shporer, A., \& Brown, T. 2011, The Astrophysical Journal, 733, 30, \dodoi{10.1088/0004-637X/733/1/30}

\bibitem[{Siegel {et~al.}(2023)Siegel, Winn, \& Albrecht}]{Siegel2023}
Siegel, J.~C., Winn, J.~N., \& Albrecht, S.~H. 2023, The Astrophysical Journal Letters, 950, L2, \dodoi{10.3847/2041-8213/acd62f}

\bibitem[{{Southworth}(2011)}]{Southworth2011}
{Southworth}, J. 2011, \mnras, 417, 2166, \dodoi{10.1111/j.1365-2966.2011.19399.x}

\bibitem[{{Stefansson} {et~al.}(2017){Stefansson}, {Mahadevan}, {Hebb}, {Wisniewski}, {Huehnerhoff}, {Morris}, {Halverson}, {Zhao}, {Wright}, {O'rourke}, {Knutson}, {Hawley}, {Kanodia}, {Li}, {Hagen}, {Liu}, {Beatty}, {Bender}, {Robertson}, {Dembicky}, {Gray}, {Ketzeback}, {McMillan}, \& {Rudyk}}]{Stefansson2017}
{Stefansson}, G., {Mahadevan}, S., {Hebb}, L., {et~al.} 2017, \apj, 848, 9, \dodoi{10.3847/1538-4357/aa88aa}

\bibitem[{Stef{\`a}nsson {et~al.}(2022)Stef{\`a}nsson, Mahadevan, Petrovich, Winn, Kanodia, Millholland, Maney, Ca{\~n}as, Wisniewski, Robertson, Ninan, Ford, Bender, Blake, Cegla, Cochran, Diddams, Dong, Endl, Fredrick, Halverson, Hearty, Hebb, Hirano, Lin, Logsdon, Lubar, McElwain, Metcalf, Monson, Rajagopal, Ramsey, Roy, Schwab, Schweiker, Terrien, \& Wright}]{rmfit_Stefansson2022}
Stef{\`a}nsson, G., Mahadevan, S., Petrovich, C., {et~al.} 2022, The Astrophysical Journal Letters, 931, L15, \dodoi{10.3847/2041-8213/ac6e3c}

\bibitem[{Storn \& Price(1997)}]{PyDE_Storn1997}
Storn, R., \& Price, K. 1997, Journal of Global Optimization, 11, 341, \dodoi{10.1023/A:1008202821328}

\bibitem[{{Tamburo} {et~al.}(2022){Tamburo}, {Muirhead}, {McCarthy}, {Hart}, {Gracia}, {Vos}, {Bardalez Gagliuffi}, {Faherty}, {Theissen}, {Agol}, {Skinner}, \& {Sagear}}]{Tamburo2022a}
{Tamburo}, P., {Muirhead}, P.~S., {McCarthy}, A.~M., {et~al.} 2022, \aj, 163, 253, \dodoi{10.3847/1538-3881/ac64aa}

\bibitem[{Tayar {et~al.}(2022)Tayar, Claytor, Huber, \& {van Saders}}]{Tayar2022}
Tayar, J., Claytor, Z.~R., Huber, D., \& {van Saders}, J. 2022, The Astrophysical Journal, 927, \dodoi{10.3847/1538-4357/ac4bbc}

\bibitem[{{Valenti} \& {Fischer}(2005)}]{Va05}
{Valenti}, J.~A., \& {Fischer}, D.~A. 2005, \apjs, 159, 141, \dodoi{10.1086/430500}

\bibitem[{Vanderburg \& Johnson(2014)}]{Keplerspline_Vanderburg2014}
Vanderburg, A., \& Johnson, J.~A. 2014, Publications of the Astronomical Society of the Pacific, 126, 948, \dodoi{10.1086/678764}

\bibitem[{{Winn} {et~al.}(2010){Winn}, {Fabrycky}, {Albrecht}, \& {Johnson}}]{Winn2010b}
{Winn}, J.~N., {Fabrycky}, D., {Albrecht}, S., \& {Johnson}, J.~A. 2010, \apjl, 718, L145, \dodoi{10.1088/2041-8205/718/2/L145}

\bibitem[{{Winn} {et~al.}(2009){Winn}, {Johnson}, {Fabrycky}, {Howard}, {Marcy}, {Narita}, {Crossfield}, {Suto}, {Turner}, {Esquerdo}, \& {Holman}}]{Winn2009b}
{Winn}, J.~N., {Johnson}, J.~A., {Fabrycky}, D., {et~al.} 2009, \apj, 700, 302, \dodoi{10.1088/0004-637X/700/1/302}

\bibitem[{{Yee} {et~al.}(2023){Yee}, {Winn}, {Hartman}, {Bouma}, {Zhou}, {Quinn}, {Latham}, {Bieryla}, {Rodriguez}, {Collins}, {Alfaro}, {Barkaoui}, {Beard}, {Belinski}, {Benkhaldoun}, {Benni}, {Bernacki}, {Boyle}, {Butler}, {Caldwell}, {Chontos}, {Christiansen}, {Ciardi}, {Collins}, {Conti}, {Crane}, {Daylan}, {Dressing}, {Eastman}, {Essack}, {Evans}, {Everett}, {Fajardo-Acosta}, {For{\'e}s-Toribio}, {Furlan}, {Ghachoui}, {Gillon}, {Hellier}, {Helm}, {Howard}, {Howell}, {Isaacson}, {Jehin}, {Jenkins}, {Jensen}, {Kielkopf}, {Laloum}, {Leonhardes-Barboza}, {Lewin}, {Logsdon}, {Lubin}, {Lund}, {MacDougall}, {Mann}, {Maslennikova}, {Massey}, {McLeod}, {Mu{\~n}oz}, {Newman}, {Orlov}, {Plavchan}, {Popowicz}, {Pozuelos}, {Pritchard}, {Radford}, {Reefe}, {Ricker}, {Rudat}, {Safonov}, {Schwarz}, {Schweiker}, {Scott}, {Seager}, {Shectman}, {Stockdale}, {Tan}, {Teske}, {Thomas}, {Timmermans}, {Vanderspek}, {Vermilion}, {Watanabe}, {Weiss}, {West}, {Van Zandt}, {Zejmo}, \& {Ziegler}}]{Yee2023}
{Yee}, S.~W., {Winn}, J.~N., {Hartman}, J.~D., {et~al.} 2023, \apjs, 265, 1, \dodoi{10.3847/1538-4365/aca286}

\end{thebibliography}
\bibliographystyle{aasjournal}



\end{document}